\documentstyle[prl,aps,twoside,floats,epsf]{revtex}

\begin{document}
\draft

\twocolumn[\hsize\textwidth\columnwidth\hsize\csname
@twocolumnfalse\endcsname
\title{Ferroelectric Phase Transitions in Films with Depletion Charge}
\author{A.M. Bratkovsky$^{*}$}
\address{Hewlett-Packard Laboratories, 3500 Deer Creek Road, MS
26U-12, Palo Alto,\\ 
California 94304-1392}
\author{A.P. Levanyuk$^{\dagger}$}
\address{Departamento de F\'{i}sica de la Materia\ Condensada, C-III,
Universidad\\ 
Aut\'{o}noma de Madrid, 28049 Madrid, Spain}
\date{ August 4, 2000  }
\maketitle

\begin{abstract}
We consider ferroelectric phase transitions in both short-circuited 
and biased ferroelectric-semiconductor films with a space (depletion)
charge which leads to some unusual behavior.
It is shown that in the presence of the charge the polarization
separates into `switchable' and `non-switchable' parts.  
The electric field, appearing due to the space charge, does not
wash out the phase transition, which remains second order but takes
place at somewhat reduced temperature. At the same time, it leads to a
suppression of the ferroelectricity in a near-electrode layer. This
conclusion is valid for materials with both second and first order
phase transitions in pure bulk samples. Influence of the depletion 
charge on thermodynamic  coercive field reduces mainly to the lowering
of the phase 
transition temperature, and its effect is negligible.
The depletion charge can, however,
facilitate an appearance of the domain structure which would be
detrimental for device performance (fatigue). We discuss some issues
of conceptual character, which are generally known but were
overlooked in previous works. The present results have general
implications for small systems with depletion charge. 

\end{abstract}


\vskip 2pc ] 

\narrowtext

\section{Introduction}

Ferroelectric perovskite materials used in memory applications usually
behave as semiconductors\cite{waser}. It is expected, therefore, that at the
metal-ferroelectric contact a depletion zone and the corresponding
(depletion) charge distribution is formed, similar to situation with usual
semiconductor-metal contacts. The corresponding electronic structure and
related effects were extensively discussed for layered metal-semiconducting
ferroelectric systems \cite
{ivanchik61,vul70,batra73,chensky82,blom94,shih94,wat98}. The depletion
effects are especially important in switching devices \cite{blom94}. The
semiconducting behavior of some common ferroelectrics (PZT) was exploited in
a series of papers \cite{tagpaw,pawtag,tagant} to account for some specific
features of switching, fatigue, and rejuvenation phenomena in ferroelectric
thin films. 

The idea of possible importance of the depletion charge for the properties
of thin ferroelectric films seems fairly natural. However, as will be shown
below, ferroelectric properties in those systems are very peculiar. The
consistent treatment demonstrates, in particular, that the effect of the
depletion charge on thermodynamic coercive field is minute. This is a
consequence of long-range Coulomb (depolarizing) field which appears
simultaneously with the inhomogeneous polarization. The depolarizing field
makes the problem of finding the polarization field and the point of phase
transition non-local, and the local, and to a considerable extent, global
dielectric response {\em rigid}. The effects of the depolarizing field were
apparently neglected by the previous authors, Refs.~\cite
{tagpaw,pawtag,tagant}. Therefore, their general relevance is questionable\
because, as is shown below, the depolarizing field plays crucial role in
determining the dielectric response and the very character of the phase
transition. The depolarizing field leads to some important phenomena, like
lowering of the critical temperature, $T_{c}$, possible
monodomain-polydomain transition, etc.

In the present paper we describe the properties of ferroelectric films with
the depletion charge in terms of the phenomenological
Landau-Ginzburg-Devonshire (LGD) theory of ferroelectricity (see, e.g. \cite
{strukov}). We restrict ourselves to the case of a ferroelectric film with a
polar axis perpendicular to the film plane. It is assumed that the charge
distribution remains the same both in the paraelectric and the ferroelectric
phases.

The key point in our treatment is that the polarization is divided into
built-in (non-switchable) and ferroelectric (switchable) parts. Within the
paper this division is mathematically strict: the built-in polarization, $%
P_{b}({\bf r})$, corresponds to an extremum of the LGD thermodynamic
potential. The extremum corresponds to a minimum in the paraelectric and a
maximum in the ferroelectric phase (the latter being similar to a zero
polarization solution for a pure short-circuited crystal). We shall call the
ferroelectric polarization the difference 
\begin{equation}
P_{f}=P-P_{b}.  \label{twopol}
\end{equation}
The ferroelectric polarization is {\em switchable}: there always are two
equivalent solutions for $P_{f}({\bf r})$ in the ferroelectric phase. In the
case of a non-zero bias we include the change of the polarization due to the
bias into $P_{f},$ whereas the built-in polarization remains unchanged.

It is convenient to start with a short-circuited film of a {\it linear}{\bf %
\ }dielectric with a homogeneous density of the space charge, $\rho $. The
electric displacement $D$ in the film is given by the electrostatic equation 
{\rm div}$\vec{D}=\rho ,$ or simply 
\begin{equation}
dD/dz=\rho =\text{const},  \label{maxwell}
\end{equation}
where $z$ is the coordinate perpendicular to the film plane. Taking into
account that the voltage across the short-circuited film is zero, one finds: 
\begin{equation}
D=(z-d/2)\rho .  \label{displshort}
\end{equation}
{}From

\begin{equation}
D=\varepsilon _{0}E+P,  \label{displ}
\end{equation}
and

\begin{equation}
D=\varepsilon \varepsilon _{0}E,  \label{displlin}
\end{equation}
one obtains for the polarization $P$:

\begin{equation}
P=\frac{\varepsilon -1}{\varepsilon }D.  \label{polarviaD}
\end{equation}
Now we shall consider a ferroelectric in a paraelectric state. We shall
neglect, for a moment, the nonlinearities and make some qualitative remarks.
It follows from Eq.~(\ref{polarviaD}) that $P\approx D$ when $\varepsilon
\gg 1$. Note that the electric displacement is always determined by the
charge density, it does not depend on the dielectric constant [cf. (\ref
{maxwell})]. When the phase transition is approached, the dielectric
constant increases, but the polarization remains practically unchanged,
unlike the electric field which exhibits a strong temperature dependence. At
the Curie temperature ($\varepsilon =\infty $) the field is zero in the
linear approximation, while the polarization is almost the same as it was
far from the transition. In other words it is more appropriate to talk about
a {\em built-in polarization} rather than a built-in electric field while
considering the effects of space charge in ferroelectrics. This conclusion
is not limited, of course, to the case of homogeneous distribution of the
space charge. For a medium with a large dielectric constant the value of the
polarization is close to the value of the displacement field (\ref{polarviaD}%
). Hence, near complete compensation of the space charge (produced by
ionized donors, etc.) by the bound charges, corresponding to {\em %
inhomogeneity} of the polarization, occurs in a `soft' dielectric medium.

The difference between the ferroelectric and non-ferroelectric (built-in)
polarization can be clarified by considering effect of the bias voltage. One
finds for a linear dielectric

\begin{equation}
D=(z-d/2)\rho +\varepsilon \varepsilon _{0}U/d,  \label{displvolt}
\end{equation}
where $U$ is the bias voltage and the polarization $P$ naturally
splits into two parts, the switchable (`ferroelectric') polarization
$P_f$ and non-switchable (`built-in') polarization $P_b$:

\begin{eqnarray}
P &=&P_{f}+P_{b};  \label{polvolt} \\
P_{f} &\equiv &(\varepsilon -1)\varepsilon _{0}U/d, \\
P_{b}(z) &\equiv &(z-d/2)\rho \frac{\varepsilon -1}{\varepsilon }.
\end{eqnarray}
Thus, in the presence of both the space
charge and the bias voltage, there are two contributions to the
polarization, but only one of them `feels' the phase transition. The
built-in polarization {\em does not change }with the bias, it remains
inhomogeneous and almost insensitive to the phase transition [since $%
(\epsilon -1)/\epsilon \approx 1]$. The induced (ferroelectric) polarization 
$P_{f}=(\varepsilon -1)\varepsilon _{0}U/d\equiv $ $(\varepsilon
-1)\varepsilon _{0}E_{\text{ext}}$ is homogeneous. The corresponding
susceptibility $\chi _{f}\equiv dP_{f}/dE_{\text{ext}}\propto
\varepsilon - 1$ diverges at the 
transition temperature in the linear approximation. We provide more details
and discussion about the division of the polarization into ferroelectric and
non-ferroelectric parts below.


The paper is organized as follows. In Sec.~II we analyze the problem of
phase transition in a short-circuited film with homogeneous space charge. We
assume that ferroelectric material is having a second-order phase transition
in the bulk. We shall show that the phase transition in the film with
depletion charge is inhomogeneous, with both `global' and `local' phase
transition temperatures being lower in the film than in the bulk. Thus, the
presence of the space charge will be shown to somewhat suppress the
ferroelectricity.

To treat the ferroelectric phase, in Sec.~III we consider a special exactly
solvable case of the space charge distribution, when all the charges are
located at the central plane of the film. We find that the presence of the
depletion charge leads to certain lowering of the critical temperature. The
corresponding lowering of the thermodynamic coercive field is shown to be
minute. Therefore, the possibility of a large effect of the built-in field
on switching in monodomain samples, asserted in Refs.~\cite
{tagpaw,pawtag,tagant}, is rather questionable, especially since they have
not considered the actual nucleation processes. Other types of the charge
distribution are discussed qualitatively. In Sec.~IV we show that in
materials with first order phase transitions the depletion charge also
effectively decreases the dielectric constant of paraelectric phase. We
summarize the results in Sec.~V and add there some comments of conceptual
character.

\section{Phase transition in the presence of homogeneous space charge}

\subsection{Main equations}

The nonlinear electric equation of state has a usual Landau form:

\begin{equation}
A P + B P^{3} = E,  \label{eqofstate}
\end{equation}
where $P(E)$ are the $z$-components of the polarization (electric field). We
assume, as usual:

\begin{equation}
A=\alpha (T-T_{c}),\qquad B={\rm const,}  \label{AB}
\end{equation}
where $T_{c}$ is the Curie temperature and $\alpha =$const. Together with
Eqs.(\ref{maxwell}),(\ref{displ}) and the boundary condition

\begin{equation}
\int_{0}^{d} E dz = U,  \label{shortcircuit}
\end{equation}
we now have the complete set of equations to treat the ferroelectric phase
transition in an electroded film. Note that we cannot use Eq.~(\ref
{displvolt}) now because it implies a \ linear relation between $D$ and $E$,
whereas we are interested in a proper account for the non-linear effects.

Integrating Eq.~(\ref{maxwell}), we find with the use of (\ref{displ}): 
\begin{equation}
\varepsilon _{0}E(z)+P(z)=\rho z+K,  \label{solutionmax}
\end{equation}
where $K$ is a constant which should be found from Eq.~(\ref{shortcircuit}).
We obtain: 
\begin{equation}
\varepsilon _{0}E(z)=-P(z)+(z-d/2)\rho +\frac{1}{d}\int_{0}^{d}P(z)dz+%
\varepsilon _{0}E_{\text{ext}},  \label{field}
\end{equation}
where the external field is produced by external bias voltage, 
\[
E_{\text{ext}}=U/d. 
\]
Combining this with Eq.~(\ref{eqofstate}) we get:

\begin{eqnarray}
&&(A+\varepsilon _{0}^{-1})P+BP^{3}-\frac{1}{d\varepsilon _{0}}%
\int_{0}^{d}P(z)dz  \label{maineq} \\
&=&(z-d/2)\rho /\varepsilon _{0}+E_{\text{ext}}.
\end{eqnarray}
{}From Eqs.~(\ref{maineq}),(\ref{twopol}) we obtain, up to terms $\propto
P_{b}^{2}:$ 
\begin{eqnarray}
&&(A+3BP_{b}^{2}+\varepsilon _{0}^{-1})P_{f}+3BP_{b}P_{f}^{2}+BP_{f}^{3} 
\nonumber \\
&&-\frac{1}{d\varepsilon _{0}}\int_{o}^{d}P_{f}(z)dz=E_{\text{ext}}.
\label{maineq2}
\end{eqnarray}

\subsection{Paraelectric phase}


Let us first calculate $P_{f}$ in the paraelectric phase to the linear
approximation, i.e. for small $U$. We have to solve the simple integral
equation

\begin{equation}
(A+3BP_{b}^{2}+\varepsilon _{0}^{-1})P_{f}-\frac{1}{d\varepsilon _{0}}%
\int_{o}^{d}P_{f}(z)dz=E_{\text{ext}},  \label{mainlineq}
\end{equation}
assuming that $A$ is positive. Introducing a notation

\begin{equation}
I\equiv\frac{1}{d\varepsilon _{0}}\int_{o}^{d}P_{f}(z)dz,
\label{designation}
\end{equation}
we find from (\ref{mainlineq}):

\begin{equation}
P_{f}=\frac{E_{\text{ext}}+I}{A+3BP_{b}^{2}+\varepsilon _{0}^{-1}}.
\label{pfviar}
\end{equation}
To find $I$ we substitute this expression into Eq.~(\ref{designation}) and,
as a result, find $P_{f}$ in an explicit form:

\begin{equation}
P_{f}(z)=E_{\text{ext}}\frac{1}{1-G}\frac{1}{A+3BP_{b}^{2}\left( z\right)
+\varepsilon _{0}^{-1}},  \label{pfexpl}
\end{equation}
where

\begin{equation}
G\equiv \frac{1}{d\varepsilon _{0}}\int_{o}^{d}\frac{dz}{A+3BP_{b}^{2}+%
\varepsilon _{0}^{-1}}.  \label{Gdesign}
\end{equation}
Thus:

\begin{equation}
1-G=\frac{1}{d\varepsilon _{0}}\int_{o}^{d}\frac{A+3BP_{b}^{2}}{%
A+3BP_{b}^{2}+\varepsilon _{0}^{-1}}dz\approx A+3B\langle P_{b}^{2}\rangle ,
\label{1-G}
\end{equation}
where $\langle \dots \rangle \equiv d^{-1}\int_{0}^{d}dz\ldots ,$ and we
have taken into account that typically $\varepsilon _{0}^{-1}\gg
(A,3BP_{b}^{2})$.

\begin{figure}[t]
\epsfxsize=2.4in
\epsffile{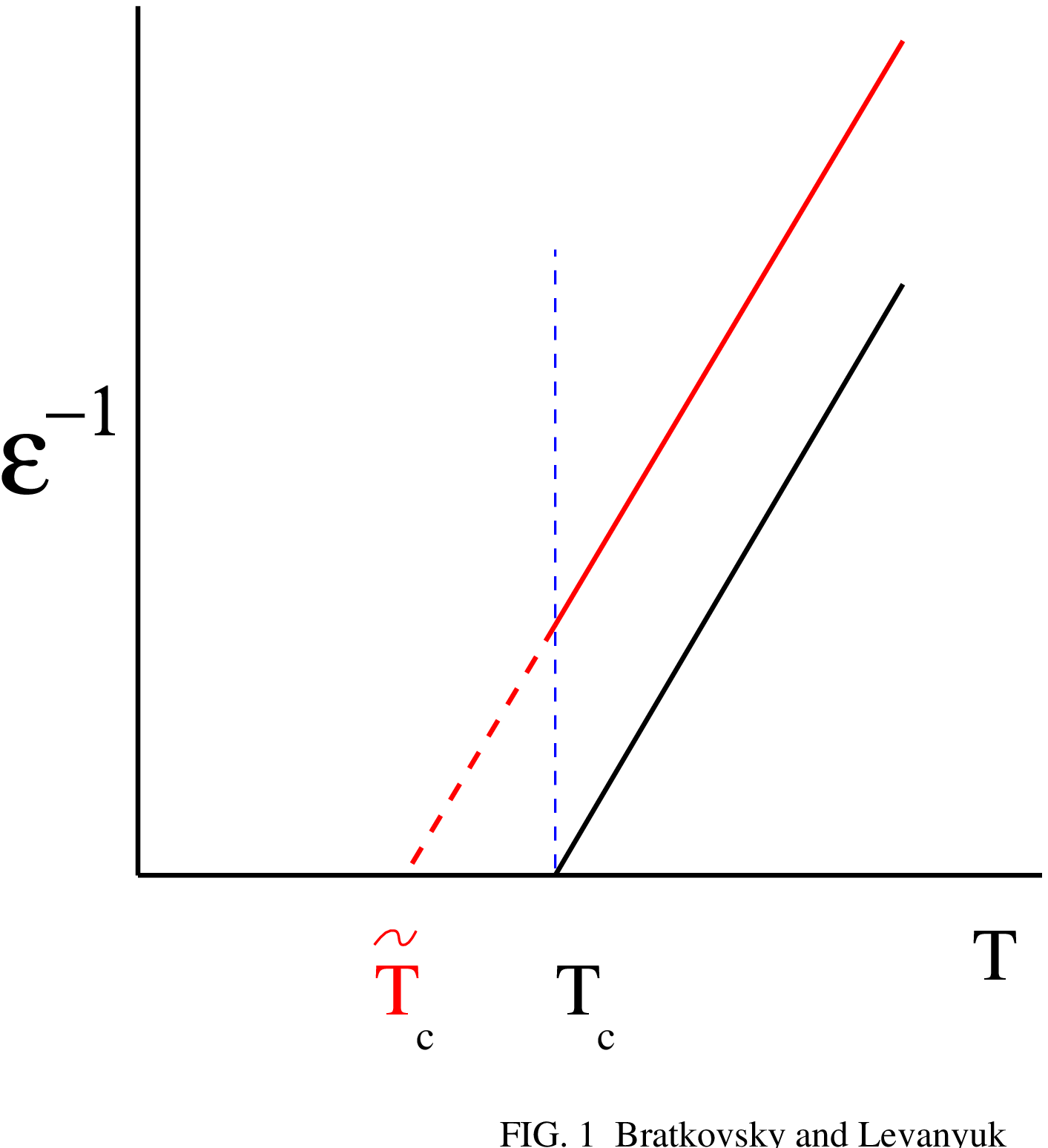 }
\vspace{.1in}
\caption{ Schematic of the dielectric constant in a pure ferroelectric film
and the film with uniform depletion charge (broken line). Depletion charge
suppresses the critical temperature from $T_{c}$ in pure sample down
to $\tilde T_c$ by $(1-100)$K (see text). }
\label{fig:1}
\end{figure}

We see, therefore, that the susceptibility $\chi_f=dP_{f}/dE_{\text{ext}%
}\propto \left( 1-G\right) ^{-1}$ diverges at the temperature $\tilde{T}%
_{c}, $ which is {\em lower} that the Curie-Weiss temperature $T_{c}$ in
pure crystal (Fig.~\ref{fig:1}), and the divergence of the dielectric
response would take place when $A=-3B\langle P_{b}^{2}\rangle $, i.e. at $T=%
\widetilde{T_{c}}:$

\begin{equation}
T_{c}-\widetilde{T_{c}}=\frac{3B\langle P_{b}^{2}\rangle }{\alpha }.
\label{difofT}
\end{equation}

Let us estimate this difference. For a displacive transition $B$ and $\alpha 
$ have normal `atomic' values\cite{strukov}. For $P_{b}\approx P_{{\rm at}}$%
, where $P_{{\rm at}}$ is the atomic polarization, $T_{c}-\widetilde{T_{c}}$
would be about $T_{{\rm at}}\sim 10^{4}-10^{5}$K. For the charged impurity
concentration $N_{d}$ one estimates $P_{b}\sim (N_{d}/N_{{\rm at}})(d/d_{%
{\rm at}})P_{{\rm at}}$ where $N_{{\rm at}}$ is the `atomic' concentration
(10$^{22}-10^{23})$cm$^{-3}$, $d_{{\rm at}}$ is the `atomic' distance\ (unit
cell length, 3-5$\AA )$ With the donor concentration $N_{d}=10^{18}$cm$^{-3}$
\cite{tagpaw}, the thickness of the totally depleted film (equal the width
of the depletion layer) would be about 2$\cdot 10^{3}\AA $, and we estimate $%
P_{b}\sim (10^{-2}-10^{-3})P_{{\rm at}}$ and $T_{c}-\widetilde{T_{c}}\sim
(1-100)$K. We see that the phase transition temperature can be reduced
substantially, i.e. the paraelectric phase can be considerably `harder' than
the same phase in the pure crystal.

We see already that there is a radical difference between the effects of \
external field (the bias voltage) and the built-in field (polarization) on
the ferroelectric transition. The former would smear out the phase
transition and shift the maximum of the dielectric response to higher
temperatures, while the built-in field does {\em not} lead to the maximum in
the dielectric response at $A>0.$

\subsection{Ferroelectric phase}

But what happens at $A<0$? This is not so easy to determine. We shall try to
answer this question, at least semi-quantitatively, but first we have to
explain what the difficulty is. It is seen from Eq.~(\ref{pfexpl}) that $%
P_{f}$ is inhomogeneous over the sample, being smaller near the electrodes.
It is worth mentioning that inhomogeneity of $P_{f}$ under the bias voltage
is different from the inhomogeneity of \ the dielectric susceptibility of
the paraelectric phase. Indeed, the `local' dielectric susceptibility is
what one would find by cutting out a small piece of the ferroelectric at a
given position in the sample, making a capacitor out of this piece, and
measuring its dielectric response. In our case we have to `fill up' the
capacitor with a medium with a given local values of $P_{b}$: i.e. it is the
built-in polarization that makes the ferroelectric response of the film
effectively inhomogeneous. If $P_{b}$ were constant, the Eq.~(\ref{mainlineq}%
) would have had a homogeneous solution:

\begin{equation}
(A+3BP_{b}^{2})P_{f}=E_{\text{ext}}.  \label{inhomsuscept}
\end{equation}
One can see that the local susceptibility $\chi _{loc}(z)\equiv
dP_{f}/dE_{ext}=\left( A+3BP_{b}^{2}\right) ^{-1}$ is {\em more}
inhomogeneous than the polarization Eq.~(\ref{pfexpl}): one has to compare $%
3BP_{b}^{2}(z)$ in the case of $\chi _{loc}$, where $A$ is small, with $%
A+\varepsilon _{0}^{-1}$ $\approx $ $\varepsilon _{0}^{-1}>>A$ in the latter
case. The {\em smaller} inhomogeneity of the polarization $P_{f}$ in
relatively {\em more} inhomogeneous dielectric medium $\left( \chi
_{loc}\right) $ is due to the mutual Coulomb interaction of the bound
charges (`longitudinal' inhomogeneity of the polarization). Consequently,
the system tries to lower this energy by reducing the
inhomogeneity.Therefore, the phase transition takes place in a fairly
strongly inhomogeneous medium. \ There is also another inhomogeneity, which
is described by the second term on the left hand side of the Eq.~(\ref
{maineq2}). It is expected to be important in the ferroelectric phase.

\begin{figure}[t]
\epsfxsize=2.4in
\epsffile{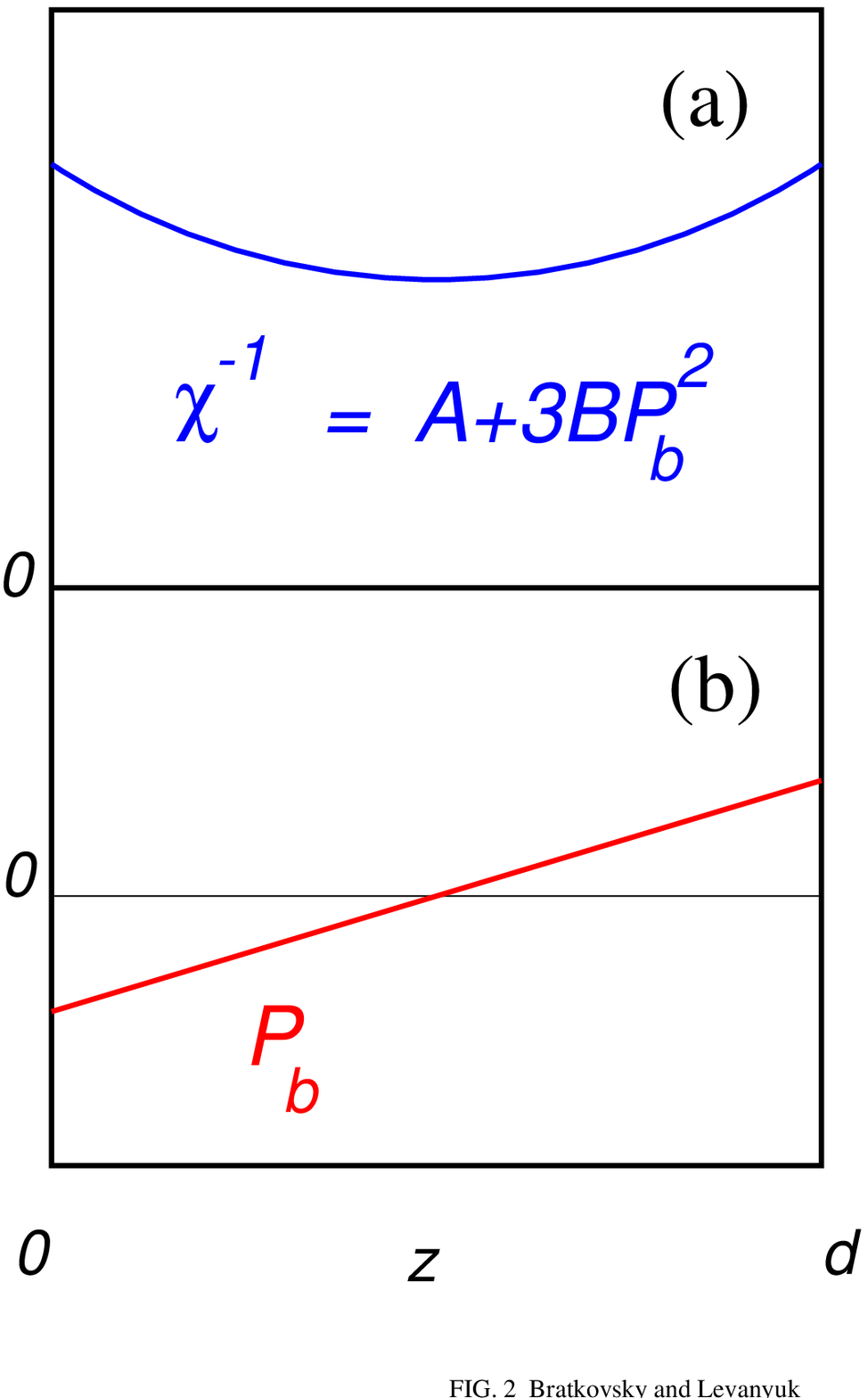 }
\vspace{.1in}
\caption{ Schematic of the dielectric response $\protect\chi_f
^{-1}=dE_{\rm ext}/dP$
which becomes inhomogeneous in the presence of the depletion charge (a), and
the built-in polarization $P_{b}$ (b).}
\label{fig:2}
\end{figure}

In the center of the film, where $P_{b}=0$, the local susceptibility is the
same as in pure crystal but at other points of the crystal it is `harder'
(meaning that $\chi _{loc}$ is lower), the hardest region being close to the
electrodes (Fig.~\ref{fig:2}). If not for long-range Coulomb interactions,
this would means that the ferroelectric phase transition would occur at the
critical temperature for pure sample, $T_{c}$. However, the Coulomb
interaction suppresses the phase transition across the system, ans $T_{c}$
reduces down to $\tilde{T}_{c},$ Fig.~1.

\begin{figure}[t]
\epsfxsize=2.6in
\epsffile{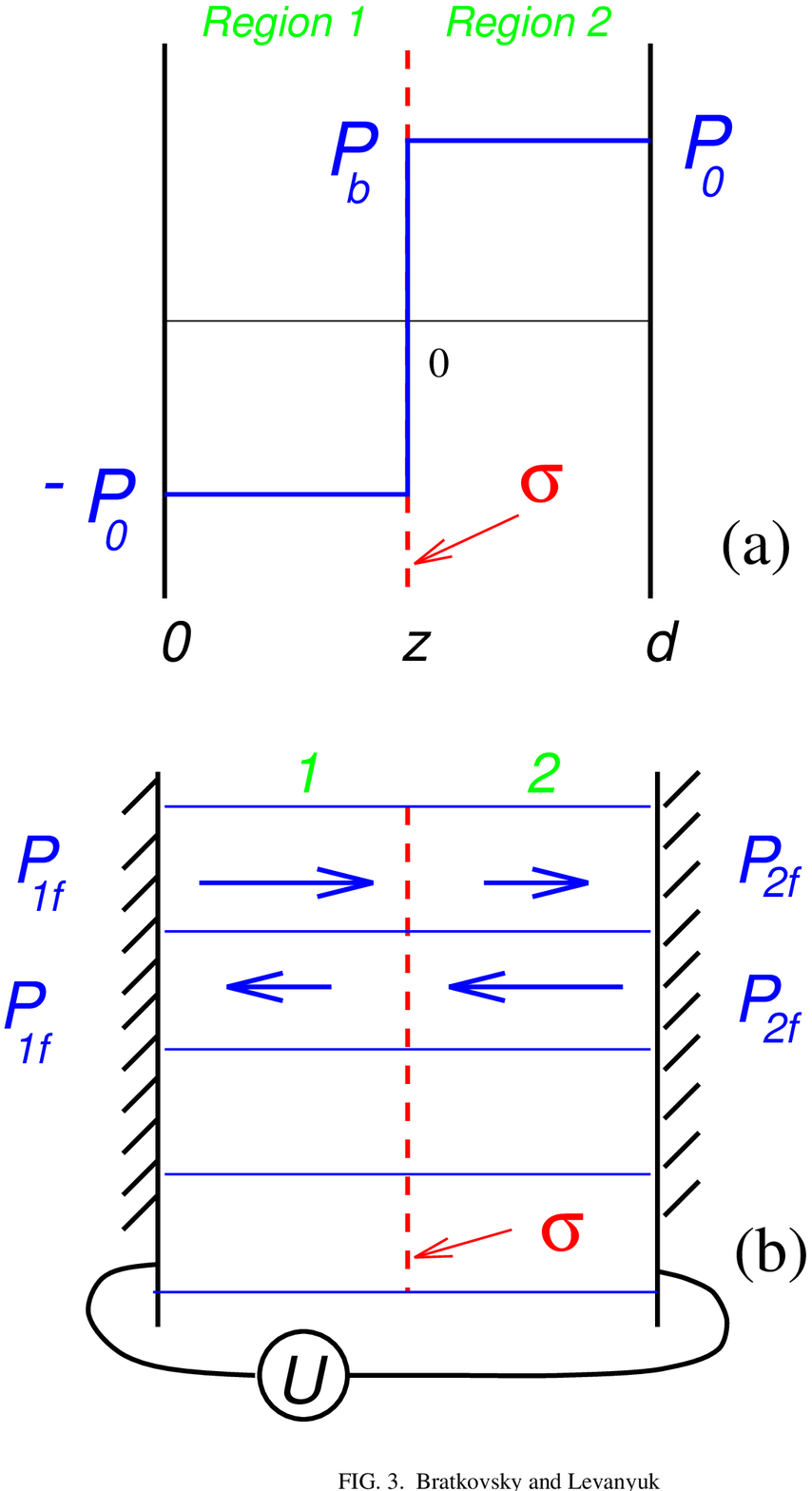 }
\vspace{.1in}
\caption{ (a) Schematic of the exactly solvable case when the charge 
(with surface density $\sigma $) is located on a plane in the middle of a
ferroelectric film under bias voltage $U$. The built-in polarization $P_{b}$
takes the values $-(+)P_{0}$, where $P_{0}=\sigma /2$, in the left
(right) halves of the film (regions 1 and 2, respectively). (b) Possible
domain structure with ferroelectric (switchable) polarization which has
different magnitude in the left and right halves of the film. }
\label{fig:3}
\end{figure}

Since the problem of the film with depletion charge proves to be fairly
complicated, it is instructive to consider first some special cases, where
the treatment is easier, and get a feel of the relevant effects. The most
natural way to simplify the problem is to replace the continuous
distribution of the built-in polarization by a stepwise distribution. The
simplest problem of that type would be a charge located at a plane in the
middle of the film with a constant charge density $\sigma $, Fig.~\ref{fig:3}%
. In this case the built-in polarization has opposite signs but the same
absolute value in the two halves of the film, so that the film is
homogeneously `hard'. It is not completely homogeneous though, there is the
inhomogeneity due to the second term on the left hand side of Eq.~(\ref
{maineq2}). But now we can study the effects of two inhomogeneities
separately, beginning with the case of a `harmonic' homogeneity and a weaker
`anharmonic' inhomogeneity. We would expect that in this case the phase
transition occurs into a {\em monodomain} state and show below that this is
a justifiable assumption.

\section{Phase transition in the film with special forms of the space charge
distribution}

\subsection{Charge located at the central plane and its effect on the
coercive field}

We denote by $P_{1}$, $E_{1}$ and $P_{2}$, $E_{2}$ the polarization and the
field in the two halves of the short-circuited film with the external bias
voltage $U$ (Fig.~\ref{fig:3}). We then have:

\begin{equation}
E_{1}+E_{2}=\frac{2U}{d}\equiv 2E_{\text{ext}},  \label{shortcirc}
\end{equation}
\begin{equation}
\varepsilon _{0}E_{2}+P_{2}-\varepsilon _{0}E_{1}-P_{1}=\sigma ,
\label{polintwohalves}
\end{equation}
\begin{equation}
AP_{1\left( 2\right) }+BP_{1\left( 2\right) }^{3}=E_{1\left( 2\right) }.
\label{eqofstatep12}
\end{equation}
Finding $E_{1}$, $E_{2}$ from the first two equations, one can present Eq.~(%
\ref{eqofstatep12}) in the form:

\begin{equation}
\left( A+\frac{1}{2\varepsilon _{0}}\right) P_{1}+BP_{1}^{3}-\frac{P_{2}}{%
2\varepsilon _{0}}=E_{\text{ext}}-\frac{\sigma }{2\varepsilon _{0}},
\label{eqofstatep1}
\end{equation}
\begin{equation}
\left( A+\frac{1}{2\varepsilon _{0}}\right) P_{2}+BP_{2}^{3}-\frac{P_{1}}{%
2\varepsilon _{0}}=E_{\text{ext}}+\frac{\sigma }{2\varepsilon _{0}}.
\label{eqofstatep2}
\end{equation}
By summing up these two equations we get:

\begin{equation}
(P_{1}+P_{2})\left[ A+B(P_{1}^{2}-P_{1}P_{2}+P_{2}^{2})\right] =2E_{\text{ext%
}}.  \label{p1+p2}
\end{equation}
Obviously, for the case of unbiased sample ($E_{{\rm ext}}=U=0$) there is
always a solution $P_{1}=-P_{2}\equiv -P_{0}$ (Fig.~\ref{fig:3}) where $%
P_{0} $ is given by

\begin{equation}
(A+\varepsilon _{0}^{-1})P_{0}+BP_{0}^{3}=\sigma /2\varepsilon _{0}.
\label{p0}
\end{equation}
Since $P_{0}$ is much smaller than the saturation polarization $P_{s}$, and
accounting for the fact that in ferroelectrics $|A|\ll \varepsilon _{0}^{-1}$
(it reads as $|A|\ll 4\pi $ in CGS units) the solution is 
\begin{equation}
P_{0}=\sigma /2.  \label{eq:P0sig}
\end{equation}
Indeed, the estimate for a typical PZT film (see below) gives $P_{0}=2.5\mu $%
Ccm$^{-2}$, whereas the saturation polarization $P_{s}=40\mu $Ccm$^{-2}$. We
shall consider this solution as the {\em built-in} polarization and present
the other solutions in the form:

\begin{equation}
P_{1\left( 2\right) }=-(+)P_{0}+P_{1\left( 2\right) f},  \label{twopol1}
\end{equation}
where $P_{1(2)f}$ is the `ferroelectric' polarization.
{}From Eqs.~(\ref
{eqofstatep1}),(\ref{eqofstatep2}) we have:

\begin{eqnarray}
&&\left( A+3BP_{0}^{2}+\frac{1}{2\varepsilon _{0}}\right) P_{1f}\nonumber\\
&-&\frac{P_{2f}}{%
2\varepsilon _{0}}-3BP_{0}P_{1f}^{2}+BP_{1f}^{3} =E_{\text{ext}}, \\
&&\left( A+3BP_{0}^{2}+\frac{1}{2\varepsilon _{0}}\right) P_{2f}\nonumber\\
&-&\frac{P_{1f}}{%
2\varepsilon _{0}}+3BP_{0}P_{2f}^{2}+BP_{2f}^{3} =E_{\text{ext}}.
\end{eqnarray}
These equations can be considered as obtained by minimization of the
thermodynamic potential: 
\begin{eqnarray}
F &=&\frac{1}{2}(\widetilde{A}+\frac{1}{2\varepsilon _{0}}%
)(P_{1f}^{2}+P_{2f}^{2})-\frac{P_{1f}P_{2f}}{2\varepsilon _{0}}  \nonumber \\
&&-BP_{0}\left( P_{1f}^{3}-P_{2f}^{3}\right) +\frac{B}{4}%
(P_{1f}^{4}+P_{2f}^{4})  \nonumber \\
&&-E_{\text{ext}}\left( P_{1f}+P_{2f}\right) ,
\end{eqnarray}
where we have introduced 
\begin{equation}
\widetilde{A}=A+3BP_{0}^{2}.  \label{eq:Atilde}
\end{equation}

It is instructive to go over to new variables, the average switchable
polarization and the difference between the switchable polarizations in both
halves of the film:

\begin{equation}
\bar{P}\equiv\frac{P_{1f}+P_{2f}}{2},\qquad \Delta P\equiv P_{1f}-P_{2f}.
\label{QviaP}
\end{equation}
In these variables the thermodynamic potential takes the form:

\begin{eqnarray}
F &=&\widetilde{A}\bar{P}^{2}+\frac{B}{2}\bar{P}^{4}-2E_{\text{ext}}\bar{P} 
\nonumber \\
&&+\frac{\widetilde{A}+\varepsilon _{0}^{-1}}{4}\Delta P^{2}-3BP_{0}\bar{P}%
^{2}\Delta P  \nonumber \\
&&-\frac{1}{2^{5/2}}BP_{0}\Delta P^{3}+\frac{B}{32}\Delta P^{4}+\frac{3}{4}B%
\bar{P}^{2}\Delta P^{2}.  \label{FQ1Q2}
\end{eqnarray}
The stability of the paraelectric phase ($P_{1f}=P_{2f}=0)$ will be lost at $%
\widetilde{A}\equiv A+3BP_{0}^{2}=0$, therefore, the parameter $\bar{P}$ is
`critical' at the phase transition while $\Delta P$ is `non-critical', since
the coefficient before $\Delta P^{2}$ does not go to zero. Now it is evident
that the phase transition we discuss is second order, since there is no
cubic term in $\bar{P}$. The external bias voltage couples only to $\bar{P}$%
, so that the phase transition for the order parameter $\bar{P}$ will be
smeared out by external field, whereas it has {\em no effect} on the
built-in polarization $P_{0}$. This highlights the qualitatively different
response of the system to external and built-in field due to space charge.

We see from the third and fourth terms in $F$ (\ref{FQ1Q2}) that the
equilibrium value of $\Delta P$ is proportional to $\bar{P}^2$ and,
therefore, the last three terms in (\ref{FQ1Q2}) would implicitly contain
terms higher order in $\bar{P}$, like $\bar{P}^{6}$. This would be
inconsistent with the form of the Landau functional that we used initially.
Therefore, the last three terms in (\ref{FQ1Q2})\ should be omitted.

Now we are in a position to find the ferroelectric polarization in the
ferroelectric phase. The equilibrium values $\bar{P}$, $\Delta P$ at $%
\widetilde{A}<0$ are:

\begin{eqnarray}
\bar{P}^{2}&\equiv & P_s^2 = -\frac{\widetilde{A}}{B},  \nonumber \\
\Delta P &=& 6\varepsilon _{0}BP_{0}\bar{P}^{2}=-6\varepsilon _{0}\widetilde{%
A}P_{0}.  \label{Q1eQ2e}
\end{eqnarray}
These formulas are approximate: we have neglected an inessential
renormalization of the coefficient $B$ due to the coupling given by the
fifth term on the right hand side in Eq.~(\ref{FQ1Q2}), and we have taken
into account that $\mid \widetilde{A}\mid <<\varepsilon _{0}^{-1}$. Then the
solution for spontaneous ferroelectric polarization is:

\begin{eqnarray}
\bar{P}_{1f} &=&\pm\sqrt{\frac{|\widetilde{A}|}{B}}+3\varepsilon _{0}|%
\widetilde{A}|P_{0}=\pm P_{s}+\frac{3}{2}\varepsilon _{0}|\widetilde{A}%
|\sigma ,  \label{P1feP2fe} \\
\bar{P}_{2f} &=&\pm\sqrt{\frac{|\widetilde{A}|}{B}}-3\varepsilon _{0}|%
\widetilde{A}|P_{0}=\pm P_{s}-\frac{3}{2}\varepsilon _{0}|\widetilde{A}%
|\sigma,
\end{eqnarray}
where +(-) sign corresponds to right- (left-) ward directed polarization in
the domains, Fig.~3(b). We see that in the half of the film where the
ferroelectric polarization has the same direction as the built-in
polarization (right half in Fig.~\ref{fig:3}), its value, $P_{2f},$ is
smaller than in another half.

\subsection{The effect of the depletion charge on a coercive field}

Now we are able to estimate the effect of the depletion charge on a coercive
field, and it happens to be very small. {}From the free energy (\ref{FQ1Q2})
we see that the dependence of the average switchable polarization $\bar{P}$
on $E_{\text{ext}}$ has a form of a hysteresis loop 
\begin{equation}
\widetilde{A}\bar{P}+\widetilde{B}\bar{P}^{3}=E_{\text{ext}}.
\label{eq:loop}
\end{equation}
Here we accounted for the fact that the fifth term in the free energy (\ref
{FQ1Q2}) is actually $\propto \bar{P}^4$ because of (\ref{Q1eQ2e}), and this
is equivalent to the renormalization of the coefficient $B \rightarrow 
\widetilde{B}\equiv B (1-18\varepsilon_0 B P_0^2)$.

\begin{figure}[t]
\epsfxsize=3.5in
\epsffile{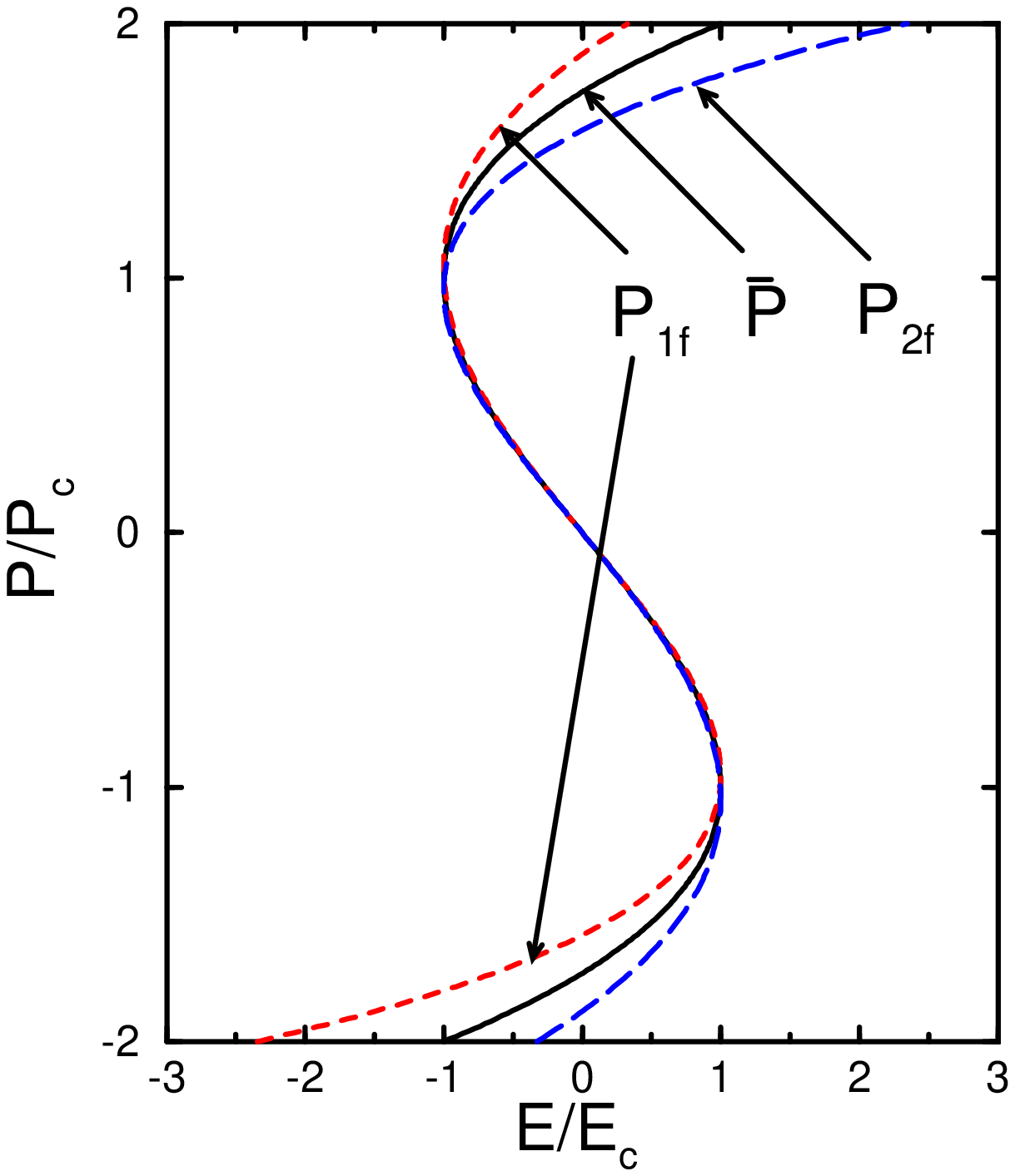 }
\vspace{.1in}
\caption{ Schematic of the hysteresis loop $P=P(E_{\rm ext})$ of the
film with the charge at 
the central plane, Fig.~3. Note that the thermodynamic coercive field
$E_c$ is 
{\em exactly the same} in both halves of the film (as seen from the
hysteresis loops for $P_{1f}$ and
$P_{2f}$). The polarization is normalized to $P_c=P(E_c)$. }
\label{fig:hloop}
\end{figure}

The `thermodynamic' coercive field for the hysteresis loop $\bar{P}=\bar{P}%
(E_{\text{ext}})$ is 
\begin{equation}
E_{c}=\frac{2\left( - \widetilde{A}\right) ^{3/2}}{3^{3/2}\widetilde{B}^{1/2}%
}.  \label{eq:Ec0}
\end{equation}
One can easily find the hysteresis loops for $P_{1f}$ and $P_{2f}$ from
Eqs.~(\ref{QviaP}),(\ref{Q1eQ2e}): 
\begin{eqnarray}
P_{1f}& =& \bar{P}(E_{{\rm ext}}) + 3\varepsilon_0 B P_0\bar{P}^2(E_{{\rm ext%
}})  \nonumber \\
P_{2f}& =& \bar{P}(E_{{\rm ext}}) - 3\varepsilon_0 B P_0\bar{P}^2(E_{{\rm ext%
}})  \label{eq:hloops}
\end{eqnarray}
The hysteresis loops are interesting since the thermodynamic coercive field
is {\em exactly the same} in both halves of the film, Fig.~\ref{fig:hloop},
in spite of that difference in {\em absolute values} of the polarizations.

Since in the ferroelectric film with depletion charge $|\widetilde{A}|<|A|,$
the coercive field there is smaller than the one, $E_{c0}$, in a sample
without the space charge. The physical reason for this reduction is the
lowering of the transition temperature at the phase transition point, where
the coercive field is zero. Thus, one can expect that the switching would be
somewhat easier in the material with the depletion charge in comparison with
a pure material. This observation is only suggestive since the actual
switching occurs at much lower fields and proceeds by nucleation and growth
processes, whereas the `thermodynamic' coercive field $E_{c}$ refers to
thermodynamic limits of stability of the homogeneous polarization in the
external field.

Let us estimate the change in the coercive field for typical thin PZT\ film
not very close to the transition, $|A|\gg 3BP_{0}^{2}$, from (\ref{eq:Ec0}) 
\begin{equation}
E_{c}=E_{c0}-\Delta E_{c},
\end{equation}
where 
\begin{eqnarray}
E_{c0} &=&\frac{2|A|^{3/2}}{3^{3/2}B^{1/2}},  \nonumber \\
\Delta E_{c} &=&-\frac{9}{2}E_{c0}{\frac{P_{0}^{2}}{P_{s}^{2}}}%
(1-2\varepsilon _{0}|A|)\approx -\frac{9}{2}E_{c0}{\frac{P_{0}^{2}}{P_{s}^{2}%
}},  \label{eq:DEc0}
\end{eqnarray}
where $E_{c0}$ is the coercive field in the pure material, and we again used
the fact that $|A|\varepsilon _{0}\ll 1$. We now consider the PZT film of
the thickness $L=300$ nm, the saturation polarization $P_{s}=40$ $\mu $C/cm$%
^{2}$, and the donor concentration $N_{d}=10^{18}$cm$^{-3}$, discussed
previously \cite{tagpaw}. With these parameters $\sigma =N_{d}L=5\mu $C/cm$%
^{2}$, $P_{0}=\sigma /2=2.5\mu $C/cm$^{2}$. We then find from (\ref{eq:DEc0}%
) 
\[
\frac{\Delta E_{c}}{E_{c}}=\frac{9}{2}\left( \frac{P_{0}}{P_{s}}\right) ^{2}=%
\frac{9}{8}\left( \frac{\sigma }{P_{s}}\right) ^{2}=0.02
\]
Obviously, the effect quickly gets {\em even smaller} for {\em thinner}
films of practical interest, $\Delta E_{c}\propto \sigma ^{2}\propto L^{2}$.

Thus, we see that the effect of the depletion charge on the coercive field
exists, but it is negligible. It has been speculated that the 'experimental'
coercive field for nucleation $E_{cn}$ should be reduced by exactly the
value of the built-in field $E_{b}$, $E_{cn}=E_{{\rm ext}}+E_{b}$ \cite
{tagpaw,pawtag,tagant}. Since these authors have estimated
$E_{b}\approx 100$kV/cm,
whereas the observed coercive field was $E_{c}\approx
80$kV/cm\cite{tagpaw}, one 
would conclude that 
the coercive field has been suppressed by more than a half by the built-in
field. However, this result and the whole previous analysis are rather
questionable, since no actual nucleation processes have been
considered in Refs.~\cite{tagpaw,pawtag,tagant}.

\subsection{On the possible domain structure}

One sees from Eq.~(\ref{P1feP2fe}) that there is a discontinuity of the
ferroelectric polarization at the central plane. It creates depolarizing
field, which can, in principle, be reduced by formation of a domain
structure [Fig.~\ref{fig:3}(b)]. To see if it really takes place we have to
use the LGD thermodynamic potential with the energy of the electric field
included explicitly:

\begin{equation}
F_{LGD}=\int dV\left[ \frac{A}{2}P^{2}+\frac{B}{4}P^{4}+\frac{\kappa }{2}%
\left( \frac{dP}{dx}\right) ^{2}+\frac{\varepsilon _{0}E^{2}}{2}\right] ,
\end{equation}
where the gradient term is added because we would like to consider a
possibility of domain structure formation with the domain walls
perpendicular to $x$ direction$,$ and this term gives rise to a surface
energy of the walls. Once more we divide the polarization into two parts as
in Eq.~(\ref{twopol}) and write

\begin{equation}
E=E_{b}+E_{f},  \label{twofields}
\end{equation}
with obvious notations. Then

\begin{equation}
F_{LGD}=F_{1}+F_{2}+F_{3},
\end{equation}
where

\begin{eqnarray}
F_{1} &=&\int dV\Biggl[\frac{A+3BP_{b}^{2}}{2}P_{f}^{2}+BP_{b}P_{f}^{3}+%
\frac{B}{4}P_{f}^{4}  \nonumber \\
&&+\frac{\kappa }{2}\left( \frac{dP_{f}}{dx}\right) ^{2}+\frac{\varepsilon
_{0}E_{f}^{2}}{2}\Biggr],
\end{eqnarray}
\begin{eqnarray}
F_{2} &=&\int dV\left( AP_{b}P_{f}+BP_{b}^{3}P_{f}+\varepsilon
_{0}E_{b}E_{f}\right)  \nonumber \\
&=&\int dV\left( E_{b}P_{f}+\varepsilon _{0}E_{b}E_{f}\right) =\int
dVE_{b}D_{f},
\end{eqnarray}
\begin{equation}
F_{3}=F_{3}(P_{b},E_{b}).  \label{F3}
\end{equation}
The energy $F_{2}=0$ for the problem of calculating the energy of the domain
structure periodic along the $x$ axis, and we can consider $F_{1}$ only. The
domain structure is shown in Fig.~\ref{fig:3}. We suppose that the domain
width is $a<d$, otherwise the domain structure would not form. The electric
field is created by the bound charges at the central plane of the film. This
field is approximately

\begin{equation}
\frac{1}{2\varepsilon _{0}}(\bar{P}_{1f}-\bar{P}_{2f})=\frac{3\widetilde{A}}{%
2}P_{0},  \label{thefield}
\end{equation}
and it is concentrated in a layer of width $\approx a$. The energy of this
field per unit area and unit length along $x$ direction is

\begin{equation}
\varepsilon _{0}\widetilde{A}^{2}P_{0}^{2}a,  \label{thefieldenergy}
\end{equation}
where we have omitted a numerical factor. The surface energy of the domain
walls for the same region is

\begin{equation}
\frac{\gamma d}{a},  \label{domwallenergy}
\end{equation}
where $\gamma $ is the surface energy of the domain wall. It is known to be 
\cite{strukov}

\begin{equation}
\gamma \approx \frac{\widetilde{A}^{2}}{B}\xi ,  \label{surfenergy}
\end{equation}
where $\xi $ is the domain wall width. Minimizing the total energy, one
finds the period of the suspected domain structure:

\begin{equation}
a\sim \left( d\xi \frac{\varepsilon _{0}^{-1}}{BP_{0}^{2}}\right) ^{1/2}.
\label{perioddomstr}
\end{equation}
The condition $a<d$ is fulfilled if

\begin{equation}
d>\frac{\varepsilon _{0}^{-1}}{BP_{0}^{2}}\xi \sim \frac{P_{{\rm at}}^{2}}{%
P_{0}^{2}}\xi \sim (10^{4}-10^{6})\xi .  \label{nodomstr}
\end{equation}
We have already mentioned that $P_{{\rm at}}/P_{0}\sim 10^{2}-10^{3}$, so
that the prefactor of $\xi $ is about $10^{4}-10^{6}$. Taking into account
that $\xi $ is no less than the unit cell length, and that it can easily be
an order of magnitude larger than that, it seems doubtful that the formation
of the domain structure due to discontinuity of the polarization in the
center of the film would be possible in films with $d\lesssim 10^{3}d_{{\rm %
at}}$. In any case, if it forms, then at fairly low temperatures, so that
our initial assumption that the phase transition takes place into a
monodomain state is justified. We see that the `anharmonic' inhomogeneity is
not the most important reason for the domain structure formation, and we
shall neglect it in the future.

\subsection{Charge located at two symmetric planes}

\begin{figure}[tbp]
\epsfxsize=2.6in
\epsffile{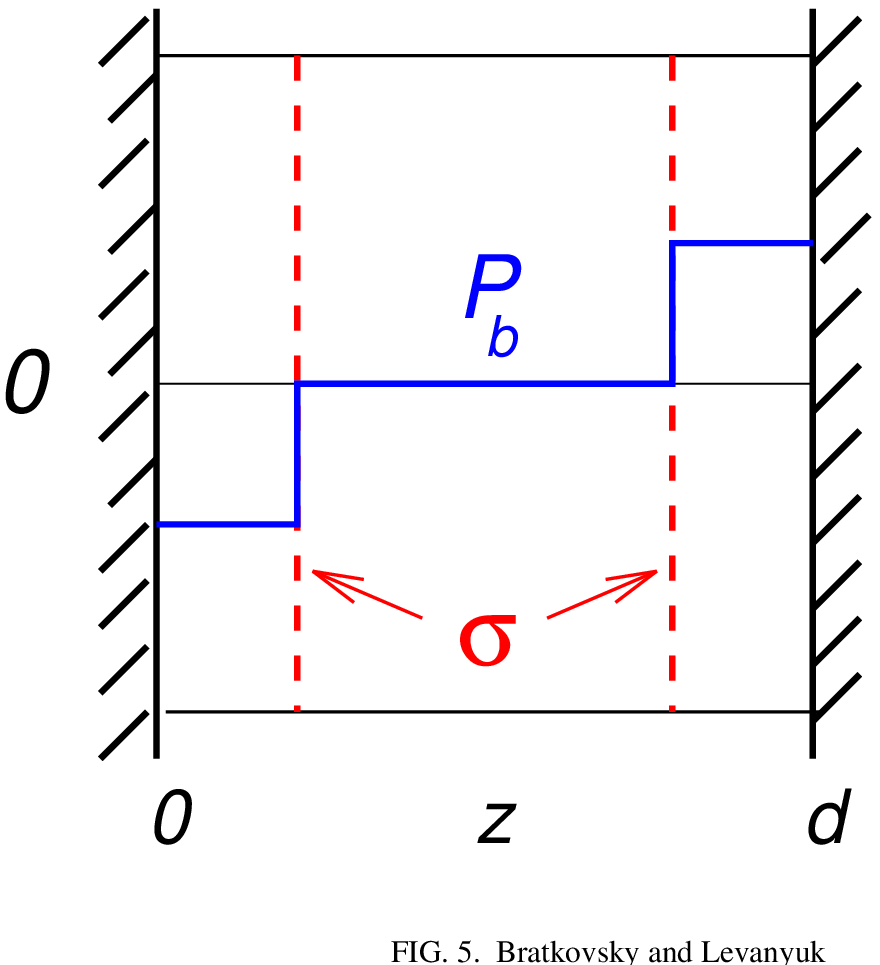 }
\vspace{.1in}
\caption{ Schematic of a ferroelectric film with two charged planes inside.
The charged planes lead to piece-wise built-in polarization $P_{b}$. }
\label{fig:2pl}
\end{figure}

A stepwise distribution of the built-in polarization, which is more like
that in Fig.~\ref{fig:2}, can be obtained with two similar symmetrically
positioned charged planes (Fig.~\ref{fig:2pl}). We have now two `hard' near
electrode regions and the central `soft' one. For $A+3BP_{0}^{2}>0$ the
system resembles a plate of a pure ferroelectric with two `passive' layers
close to the electrodes. Phase transitions in a similar system
(ferroelectric plate with vacuum gaps near the electrodes) were studied by
Chensky and Tarasenko\cite{chenski}. They found that there are transitions
to both monodomain and polydomain states depending on geometrical
parameters. Generalizing their consideration to the case of two dielectric
layers we found that the transition to the monodomain state takes place if

\begin{equation}
d_{1}<d_{1c}=\varepsilon _{1}\left( \frac{12\varepsilon _{0}\kappa }{%
\varepsilon _{\perp }}\right) ^{1/2},  \label{condition}
\end{equation}
where $d_{1}$, $\varepsilon _{1}$ are the width and the dielectric constant
of the `passive' layer, $\varepsilon _{\perp }$ is the dielectric constant
in the plane of the film (perpendicular to easy axis). In our case the value
of $d_{1c}\sim d$ can be estimated as:

\begin{equation}
d_{1}\sim \frac{P_{{\rm at}}^{2}}{P_{0}^{2}}d_{{\rm at}}\sim (10^{4}-10^{6})%
\mathop{\rm \AA }%
.  \label{d1courcase}
\end{equation}
This estimate is certainly crude but it seems likely that the phase
transition in completely depleted films proceeds into a monodomain state. We
have recently discussed a domain structure in a ferroelectric plate with
`passive' layers\cite{bratlev}. It was found that far from the phase
transition a domain structure exists for however thin passive layer.
Therefore, one can expect that a domain structure forms at some temperature
below the phase transition in a short-circuited completely depleted
ferroelectric film, but this question should to be studied in greater detail.

\section{First order phase transition}

It might be tempting to conclude, based on Eq.~(\ref{difofT}), that while
the built-in polarization `hardens' ferroelectrics with second order
transition it `softens' those with first order transitions, since the
coefficient at $P^{4}$ in the LGD thermodynamic potential is negative in the
latter case ($B<0$). However, such a hastily conclusion is almost certainly
wrong.

The fact of the matter is that, while considering relatively `weak' first
order phase transition (otherwise the LGD theory is not applicable), one has
to take into account explicitly the interaction of the order parameter with
elastic deformations. Specific features of the elasticity in solids make the
above statement about the sign of the coefficient $B$ before $P^{4}$ fairly
vague. There are several different $P^{4}$ coefficients and some of them
remain positive for weakly first order transitions\cite{levanyuk}. To treat
the problem, one has to start with the LGD potential in the form:

\begin{eqnarray}
F &=&\frac{A}{2}P^{2}+\frac{B}{4}P^{4}+\frac{C}{6}P^{6}  \nonumber \\
&&+rP^{2}u_{ll}+\frac{K}{2}u_{ll}^{2}+\mu \left( u_{ik}-\frac{1}{3}\delta
_{ik}u_{ll}\right) ^{2},  \label{LGDfirstorder}
\end{eqnarray}
where $u_{ik}$ is the strain tensor, $r$ is the electrostriction constant, $%
K $, $\mu $ are the bulk and the shear moduli. For a bulk sample the phase
transition takes place usually in a free crystal and after minimization over
the strain components one obtains:

\begin{equation}
F=\frac{A}{2}P^{2}+\frac{\widetilde{B}}{4}P^{4}+\frac{C}{6}P^{6},
\label{LGDfirstorderen}
\end{equation}
where

\begin{equation}
\widetilde{B}=B-\frac{2r^{2}}{K}.  \label{renB}
\end{equation}
The first order of the phase transition means $\widetilde{B}<0,$ but the
coefficient $B$ itself may be positive and it is {\em indeed} positive and
large for not too `strong' first order {\it structural} transitions, because 
$2r^{2}/K$ is of `atomic' value in this case, and $\widetilde{B}$ is rather
small.

But for a phase transition in an inhomogeneous system the above procedure
does not work. One cannot minimize over the strain components because they
are not independent there\cite{landau}. For our case, one can assume that
all the displacements are perpendicular to the film plane so that the only
relevant component of the strain tensor is $u_{zz}$. Putting all other
strain components in Eq.~(\ref{LGDfirstorder}) to zero and minimizing over $%
u_{zz} $ we obtain:

\begin{equation}
F=\frac{A}{2}P^{2}+\frac{\widehat{B}}{4}P^{4}+\frac{C}{6}P^{6},
\label{LGDfilm}
\end{equation}
where

\begin{equation}
\widehat{B}=B-\frac{2r^{2}}{K+\frac{4}{3}\mu }=\widetilde{B}+\frac{2r^{2}}{K}%
\frac{\frac{4}{3}\mu }{K+\frac{4}{3}\mu }.  \label{Brenormfilm}
\end{equation}
It is worth mentioning that $K+\frac{4}{3}\mu $ is the modulus with respect
to uniaxial extension without changing the lateral size\cite{landau1}. The
last factor in Eq.~(\ref{Brenormfilm}) is usually of the order of unity and
if $\mid B\mid $ is not very large then $\widehat{B}>0$ and the results of
Sec.~II are applicable for the case considered there as well.

One might conclude that, because $\widehat{B}>0$, the first order phase
transition in a bulk sample becomes second order in the film. Such a
possibility is not excluded but several reservations are in order. Firstly,
we have assumed that only one component of the strain tensor operates. It is
a natural assumption for an infinite film but the real films are not
infinite. Secondly, even for an infinite film there are various
possibilities for the phase transition: it could, for instance, transform
into a mixture of homogeneous and inhomogeneous (two phase) state. Which of
these possibilities is realized in practice depends on the conditions that
we have not specified here, e.g. on the elastic properties of the substrate.
The nature of first order phase transitions in films is worth studying
together with taking into account the actual experimental conditions.

In conclusion of this Section we would simply note that in materials with 
{\em first} order ferroelectric transition the paraelectric phase in thin
depleted films becomes more `ferroelectrically rigid', its dielectric
constant decreases, in analogy with ferroelectrics with {\em second }order
phase transition.

\section{Comments}

We have found that the polarization in the present problem naturally
separates into `switchable' and `non-switchable' parts in the presence of
the built-in (space) charge. Obviously, only the switchable polarization may
be used as an order parameter in the problem. In our example of a system
with all the charge placed at the central plane the order parameter was $%
P-P_{0}(\vec{r})$ averaged over the sample, where $P_{0}(\vec{r})$ is the
non-switchable part of the polarization. The case of a homogeneous charge in
the ferroelectric proved to be too complicated and its consideration was
omitted. What is the problem with selecting the order parameter in the
present situation? We would like to clarify the question in this section.

One difficulty with selecting the order parameter in systems with depletion
charge is that \ these systems are inhomogeneous with respect to the phase
transition, even when the depletion charge is homogeneous. The order
parameter for a phase transition in the inhomogeneous medium is the
amplitude of a function that has a space distribution of the quantity that
plays a role of the order parameter in homogeneous medium. The function
gives the form of the `most rapidly growing fluctuation' or of the symmetry
perturbation responsible for the loss of stability of a symmetric phase.

Another difficulty stems from the polarization being an electric quantity,
intimately related to a distribution of the electric charge in a crystal. In
defining the order parameter one has to recall that in theory the notions of
the `order parameter', `phase transition', and so forth, implicitly refer to
an infinite medium. It is ambiguous, or even meaningless, however, to
consider the polarization of an infinite medium (see e.g. Ashcroft and
Mermin's book, Ch.~27). This is because the electric field in a sample is
determined by the boundary conditions (shape of the sample, presence or
absence of electrodes, bias voltage, etc.). One can avoid specifying the
boundary conditions and still remain within rigorous definitions by
considering the plane waves ${\bf P}_{k}$ of the order parameter in an
infinite medium and then taking a long wave length limit ${\bf k}\rightarrow
0$. Those waves which stiffness constant goes to zero at the phase
transition (for {\it almost} infinite wave lengths) belong to the order
parameter, otherwise they do not. In most cases this procedure is not
needed, since there is no singularity (discontinuity) at ${\bf k}=0$, so
that the values of all quantities at{\bf \ }${\bf k}=0$ and at ${\bf k}%
\rightarrow 0$ are the same. This is {\em not} valid in the case of\
polarization. Only {\em transversal} polarization waves can loose stiffness
at a critical point (i.e. the transversal mode softens to zero at $T_{c}$),
whereas the {\em longitudinal} waves, unlike the transversal ones, create a 
{\em macroscopic} electric field and, therefore, remain hard at the phase
transition. One can identify the order parameter of a (proper) ferroelectric
transition with the {\em transversal } part of the polarization, but this is
not convenient in practice. One usually considers finite size samples with
well defined boundary conditions, and it might be quite difficult (if not
impossible) to discriminate between transversal and longitudinal
polarizations. The best we can advise at this moment is to remember that
there may be a part of polarization that does not correspond to the
ferroelectric order parameter. Direct solution of the `electric' equation of
state (\ref{eqofstate}) together with Maxwell equations (\ref{maxwell}) with
account to boundary conditions\ will do the job. The simultaneous solution
of the equation of state and electrostatic equations makes it clear that,
for instance, the effects of the external field and the space charge are
qualitatively different.

Usually the complication about selecting the order parameter does not arise,
but in the present case it is convenient to discriminate between the
ferroelectric and non-ferroelectric part of polarization: what we call the
`built-in polarization' is clearly `non-ferroelectric', it is longitudinal
and practically does {\em not} change at the ferroelectric phase transition.
What we call the `ferroelectric' polarization is strictly `transversal' (in
the sense that it creates no field or, rather, this field is screened by the
charges at the electrodes) only when this polarization is homogeneous. It is
not exactly our case, so that the conceptual basis for such a division of
the polarization may easily be questioned. Therefore, we would refer to
Eqs.~(\ref{twopol}),(\ref{maineq2}) and our present comments. This division
is very helpful for the present problem, with simplifications and
complications resulting from the interaction between the two parts of the
polarization, since for ferroelectrics the `electric' equation of state is
nonlinear. The effects of this interaction were focus of the present paper.

\section{Conclusions}

We have demonstrated a peculiar nature of the phase transition in
ferroelectric films with depletion (space) charge, and specific properties
of the ferroelectric phase. The charge leads to appearance of the built-in
(frozen) polarization which is not sensitive to the ferroelectric phase
transition, and it suppresses the ferroelectricity in the near-electrode
regions. This is true of both second- and first-order phase transitions. The
lowering of the critical temperature leads to very small reduction in the
thermodynamic coercive field. The crux of the mater is that the system is
strongly affected by long-range Coulomb field, accompanying the
inhomogeneous polarization, which makes the local and, to some extent, the
global dielectric response {\em rigid }(meaning the reduced dielectric
susceptibility). The main effect is that it suppresses the critical
temperature across the whole sample. The effects of the depolarizing field
in this problem have been apparently neglected by previous authors \cite
{tagpaw,pawtag,tagant}, who also speculated about built-in field assisted
switching in ferroelectric films. Since no actual nucleation processes have
been considered in these works, we doubt those speculations have any
justification.

The unusual feature of the present situation is that the value of the
ferroelectric (switchable) part of the polarization is smaller when it is
parallel to the built-in field, and larger when it is anti-parallel. This
may facilitate a splitting of the film into domains at low temperatures,
thus playing a detrimental role in the device performance (fatigue). This
illustrates an important point that the switching behavior is defined by
both the bulk properties of the sample (electronic, ferroelectric, and
microstructural) and boundary conditions, which all should be taken into
account. The observed, i.e. hysteretic, behavior of the device may change
drastically depending on variations of any of these conditions and the
ferroelectric material itself (like the growth of the `passive' dielectric
layer at the electrode-ferroelectric interface). We expect that the behavior
discussed in the present paper is generic for systems of small size where
screening of the impurity charge becomes ineffective and depletion effect
becomes pronounced.

We appreciate useful discussions with G.A.D. Briggs, A. Gruverman, and
R.S.Williams. Extensive help by Laura Wills-Mirkarimi regarding experimental
situation was invaluable. We thank A.K. Tagantsev for conversations which
helped us to tidy up the present arguments. APL would like to acknowledge a
support and hospitality of Quantum Structures Research Initiative at
Hewlett-Packard Labs (Palo Alto) during the course of this work.


\end{document}